\documentclass[showpacs,twocolumn,pre]{revtex4}
\usepackage{eurosym}
\usepackage{graphicx}
\usepackage{epstopdf}
\usepackage{amsmath}
\usepackage{latexsym}
\usepackage{epsfig}
\usepackage{makeidx}
\usepackage{amsfonts}

\begin{document}

\title{Emergence of L\'{e}vy Walks in Systems of Interacting Individuals}
\author{Sergei Fedotov and Nickolay Korabel }
\affiliation{School of Mathematics, The University of Manchester, Manchester M13 9PL, UK}

\begin{abstract}
Recent experiments (G. Ariel, et al., Nature Comm. 6, 8396 (2015)) revealed
an intriguing behaviour of swarming bacteria: they fundamentally change
their collective motion from simple diffusion into a superdiffusive L\'{e}vy
walk dynamics. We introduce a nonlinear non-Markovian persistent random walk
model that explains the emergence of superdiffusive L\'{e}vy walks. We show
that the alignment interaction between individuals can lead to the
superdiffusive growth of the mean squared displacement and the power law
distribution of run length with infinite variance. The main result is that
the superdiffusive behaviour emerges as a nonlinear collective phenomenon,
rather than due to the standard assumption of the power law distribution of
run distances from the inception. At the same time, we find that the
repulsion/collision effects lead to the density dependent exponential
tempering of power law distributions. This qualitatively explains
experimentally observed transition from superdiffusion to the diffusion of
mussels as their density increases (M. de Jager et al., Proc. R. Soc. B 281,
20132605 (2014)).
\end{abstract}

\maketitle


The ability of living cells and bacteria to maintain a collective migration
is fundamental for many physiological and pathological processes.
Self-organized and coordinated movement in living organisms occurs during
wound repair, formation of biofilms, tumor invasion, morphogenesis, etc.\
\cite{Cell, Cell2}. Cells and bacteria have a tendency to form the cohesive
groups in which mutual interactions lead to a cooperative movement of the
whole population. A particular example of such coordinated mass migration is
a bacterial swarming, for which the movement of bacteria across a surface
involves a formation of packed clusters (\textquotedblleft swarms") \cite{Sw}%
.\ It has been found recently that the swarming bacteria can perform a L\'{e}%
vy walk \cite{Main}. The extraordinary feature of this movement is that the
emergence of a superdiffusive motility is a result of the interactions
between bacteria rather than the standard mechanism of controlling the
individual frequency of tumbling. On the other hand, it was shown that the
interactions between mussels lead to the transition from a superdiffusive L%
\'{e}vy walk to Brownian motion as the density of mussels increases \cite%
{MMM}. The shift has been explained by the frequent collisions of organisms
in a dense group.

These experiments clearly show that a \textit{nonlinear L\'{e}vy walk model }%
is needed to explain the emergence of the power-law distribution of run
distances and its tempering, since the standard models are based on the
assumption of power-law distributed run distances from the inception \cite%
{SWK,KBS,MK}. Currently the model of emerging L\'{e}vy walk in systems of
interacting individuals is not available, despite the number of publications
on \textit{L\'{e}vy movement} increased\ dramatically over the last decade
\cite{KRS,KlSo,MB,Reb,Za,Hel,Jake}. One of the reasons for increasing
interest in L\'{e}vy transport is that living organisms can use it to
accelerate pattern formation which leads to improvement of individual
fitness \cite{MMMM} and to optimize searching for sparse targets \cite%
{Be,Me,Vis2,Bart}.

In this paper we propose a \textit{nonlinear L\'{e}vy walk model} for which
the superdiffusion is an emerging collective phenomenon. Motivated by the
recent experiments \cite{Main,MMM}, we suggest the nonlinear persistent
random walk model that explain both (1) the emergence of superdiffusive
motion of bacteria within a swarm and (2) the transition from superdiffusion
to the standard Brownian motion through \textit{nonlinear tempering} of the
power law distribution of run distances. Modeling of the collective movement
of individuals such as insect swarms, bird flocks, schools of fish, etc.\
has received a lot of attention in the last decades \cite{Vicsek}. Apart
from various Lagrangian (agent-based) models, many kinetic equations have
been developed to describe interaction between individuals \cite%
{P,Lewis,FE,Fete,Car}. The crucial problem here is how to incorporate
nonlinear interaction terms into non-Markovian random walks. To implement
nonlinear effects we use the structural density approach together with a
population density dependent turning rate. This method has been used by the
authors for the analysis of subdiffusive random walks \cite{Fed,FK} and L%
\'{e}vy walks \cite{Fed2}. We take into account the interactions between
walkers on the mesoscopic level, at which the turning rate nonlocally
depends on the mean field population density (nonlinear effect) and running
time (non-Markovian effect).

\textit{Nonlinear L\'{e}vy walk model in 1-D.} We consider proliferating
individuals moving either left or right along one-dimensional space at a
constant speed $v$. The generalization for high dimensions is outlined in
the Supplementary Materials. Note that 1-D is enough to show that the
superdiffusion can be an emerging nonlinear phenomenon. The key
characteristic of such random movement is the turning rate $\mathbb{T}$,
which defines moments when individuals change their direction of movements.
We introduce the structural densities of individuals, $n_{+}(x,t,\tau )$ and
$n_{-}(x,t,\tau ),$ at location $x$ and time $t$ that move in the right
direction ($+$) or the left direction ($-$) during running time $\tau $
since the last velocity switching \cite{Alt}. The mean density of
individuals moving right, $\left( +\right) ,$ and left, $\left( -\right) ,$
are defined as%
\begin{equation}
\rho _{\pm }(x,t)=\int_{0}^{\infty }n_{\pm }(x,t,\tau )d\tau  \label{den8}
\end{equation}%
and the total density $\rho (x,t)=\rho _{+}(x,t)+\rho _{-}(x,t).$

To describe random movement of individuals with interactions, we assume that
the rate $\mathbb{T}_{\pm }$ at which individuals change their direction of
motion depends not only on running time $\tau $ (non-Markovian effect) but
also on the population densities $\rho _{+}$ and $\rho _{-}$(nonlinear
effect):
\begin{equation}
\mathbb{T}_{\pm }\left( \tau ,\rho _{+},\rho _{-}\right) =\frac{\mu _{\pm
}\left( \rho _{+},\rho _{-}\right) }{\tau _{0}+\tau }+\gamma _{\pm }\left(
\rho _{\mp }\right) ,  \label{turn}
\end{equation}%
where $\tau _{0}$ is the time parameter.\ Inverse dependence of the first
term in $\mathbb{T}_{\pm }$ on the running time $\tau $ leads to a strong
persistence of the random walk. The function $\mu _{\pm }\left( \rho
_{+},\rho _{-}\right) $ describes the alignment effects leading to
cooperative movement of individuals in one direction. If, for example, an
individual moves to the right and senses that neighbouring conspecifics move
in the same direction then the likelihood of velocity switching decreases.
The positive function $\gamma _{\pm }\left( \rho _{\mp }\right) $ takes into
account the increase in the turning rate $\mathbb{T}_{\pm }$ when the
individuals avoid the collisions with those moving in the opposite
direction. The external forces or chemotactic factors could be included
analogously to systems with subdiffusion \cite{FORCE,CHEMO}. Here we do not
consider these effects.

In this paper we model the alignment among individuals by the function:
\begin{equation}
\mu _{\pm }\left( \rho _{+},\rho _{-}\right) =\mu f\left( A_{\pm }\right) ,
\label{mu}
\end{equation}%
with a non-local density dependent function $A_{\pm }\left( x,t\right) $:
\begin{equation}
A_{\pm }=a\int_{\mathbb{R}}e^{-\frac{|z|}{l_{a}}}\left[ \alpha \rho _{\pm
}(x+z,t)-\beta \rho _{\mp }(x+z,t)\right] dz.  \label{A}
\end{equation}%
Here $\mu $ is the exponent of a power law distribution in the absence of
nonlinear interactions ($A_{\pm }=0)$, $f(x)$ is a positive and decreasing
function of $x$ with $f\left( 0\right) =1$, $a$ is the strength of
interactions and $\alpha $, $\beta $ are weight parameters. The decreasing
function $f\left( A_{\pm }\right) $ indicates that the turning rate $\mathbb{%
T}_{\pm }$ is reduced due to the presence of many conspecifics moving in the
same direction. This negative dependence plays the central role in the
transition from a standard random walk to a L\'{e}vy walk. The kernel $\exp
\left( -|z|/l_{a}\right) $ describes a strength of alignment per unit
density with the distance $|z|$, $l_{a}$ is the characteristic length of the
interaction zone. To illustrate the effect of alignment, let us consider the
case $\alpha $ $=\beta =1$ for which the non-local function $A_{+}$ for
right-moving individuals can be rewritten in terms of the flux $J=v\left(
\rho _{+}-\rho _{-}\right) .$ We can write $A_{+}=av^{-1}\int_{\mathbb{R}%
}\exp \left( -\frac{|z|}{l_{a}}\right) J(x+z,t)dz$, so the increase in the
flux $J$ leads to an increase of alignment effects and decrease of turning
rate $\mathbb{T}_{+}$. This indicates the emergence of nonlinear persistence
which, together with running time persistence, can generate superdiffusive
behaviour. Such modelling is in agreement with observation that the motion
of swarming bacteria is mostly governed by the collective flow of the
bacteria and surrounding fluids \cite{Main}. Note that the advantage of
moving in a large group in the same direction is very similar to the
\textquotedblleft peloton" phenomenon in a road bicycle race. Similar
non-local dependencies of the turning rate on the population density has
been successfully used to describe the animal spatial group patterns and
bacterial swarming in terms of the hyperbolic and kinetic models \cite%
{P,Lewis,FE,Fete,Car}.

The second term $\gamma _{\pm }>0$ in the turning rate $\mathbb{T}_{\pm }$,
Eq.\ (\ref{turn}), describes the collision/repulsion effects. We assume the
increase in the turning rate $\mathbb{T}_{\pm }$ when individuals tend to
avoid collisions with many conspecifics moving in the opposite direction:
\begin{equation}
\gamma _{\pm }\left( \rho _{\mp }\right) =r\int_{0}^{\infty }\exp \left( -%
\frac{z}{l_{r}}\right) \rho _{\mp }(x\pm z,t)dz,  \label{col}
\end{equation}%
where $l_{r}$ is the effective repulsion size, and $r$ is the strength of
repulsion. The similar repulsion rate has been well used in the hyperbolic
model \cite{Lewis} to obtain spatial patterns. In our non-Markovian model,
the role of the collision/repulsion rate $\gamma _{\pm }$ is drastically
changed. This term is responsible for the shift from the superdiffusive L%
\'{e}vy walk to diffusion as the density increases \cite{MMM}.
\begin{figure}[t]
\vspace{-70pt} \centerline{ \psfig{figure=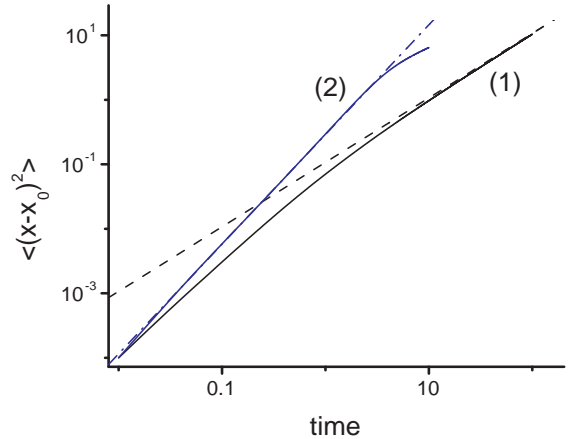,width=95mm,height=90mm} }
\caption{ Emergence of the L\'{e}vy walks when alignment dominates
repulsion. The log-log plot of the MSD (curve (2)) displays the
superdiffusive behavior with exponent $1.7$ (dashed-dotted line). We have
used $a=1.5$, $\protect\alpha =1$, $\protect\beta =0$, and $r=0$. Other
parameters are $\protect\mu =3$, $\protect\tau _{0}=0.1$, $v=1$. Without
interactions $a=0$, $r=0$ (curve (1)), the MSD grows linearly in the long
time limit (dashed line has the slope $1$). }
\label{FIG1}
\end{figure}

The nonlinear equations for the structural densities $n_{+}(x,t,\tau )$ and $%
n_{-}(x,t,\tau )$ can be written as:
\begin{equation}
\frac{\partial n_{\pm }}{\partial t}\pm v\frac{\partial n_{\pm }}{\partial x}%
+\frac{\partial n_{\pm }}{\partial \tau }=-\mathbb{T}_{\pm }\left( \tau
,\rho _{+},\rho _{-}\right) n_{\pm }.  \label{str}
\end{equation}%
We use symmetrical initial conditions for which all individuals start to
move with zero running time
\begin{equation}
n_{\pm }(x,0,\tau )=\frac{\rho (x,0)}{2}\delta (\tau ).  \label{zero}
\end{equation}%
Zero running time condition ($\tau =0)$ includes the proliferation of the
individuals:
\begin{equation}
n_{\pm }(x,t,0)=\int_{0}^{t}\left[ \mathbb{T}_{\mp }\left( \tau ,\rho
_{+},\rho _{-}\right) n_{\mp }+k\left( \rho \right) n_{\pm }\right] d\tau ,
\label{str5}
\end{equation}%
where $k\left( \rho \right) $ is the density dependent proliferation rate.
This condition corresponds to the case when newborn individuals have zero
running time. It is convenient to introduce the total turning rates defined
as
\begin{equation}
i_{\pm }(x,t)=\int_{0}^{t}\mathbb{T}_{\pm }\left( \tau ,\rho _{+},\rho
_{-}\right) n_{\pm }(x,t,\tau )d\tau .  \label{ii}
\end{equation}%
Differentiating (\ref{den8}) with respect to time $t$ together with (\ref%
{zero}) and (\ref{str5}) we derive the balance equations for the
unstructured mean densities $\rho _{+}(x,t)$ and $\rho _{-}(x,t)$:
\begin{equation}
\frac{\partial \rho _{\pm }}{\partial t}\pm v\frac{\partial \rho _{\pm }}{%
\partial x}=-i_{\pm }(x,t)+i_{\mp }(x,t)+k\left( \rho \right) \rho _{\pm }
\label{be}
\end{equation}%
(see Supplementary Materials). For L\'{e}vy walks without interactions, one
can find $i_{\pm }(x,t)=\int_{0}^{t}K(\tau )\rho _{\pm }(x\mp v\tau ,t-\tau
)d\tau $, where $K(\tau )$ is the memory kernel determined by its Laplace
transform \cite{Fed4} $\hat{K}(s)\simeq \frac{1}{T}\left( 1+As^{\mu
-1}\right) $ for $1<\mu <2$, as $s\rightarrow 0$ ($T$ is mean running time
and $A$ is a constant). For the nonlinear case, the expressions for $i_{\pm
} $ are not known. In what follows we use numerical simulations to obtain
our results.
\begin{figure}[t]
\vspace{-70pt} \centerline{ \psfig{figure=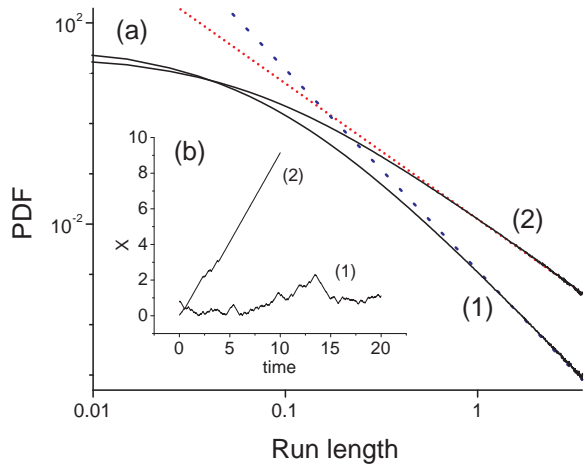,width=85mm,height=90mm} }
\caption{(a) The run length PDF of the emerging L\'{e}vy walkers (curve
(2)). Parameters are the same as in Fig.\ \protect\ref{FIG1}. The run length
PDF is a power law with the slope $-2.7$ (the dashed line), that is the
variance of the PDF is infinite. For non-interacting walkers the run length
PDF is also a power law (curve (1)), but the slope is $-\protect\mu -1=-4$
(the dotted line), so the variance is finite. (b) Typical trajectory of
interacting L\'{e}vy walkers (curve $(2)$, for better clarity we use $a=2.5$%
) is very persistent unlike the Brownian-like trajectory for a walker
without interactions (curve $(1)$). }
\label{FIG2}
\end{figure}

\textit{Emergence of superdiffusion}. To focus on the collective movement
and the underlying mechanism of the superdiffusive behaviour of the walkers,
we consider non-proliferating walkers ($k(\rho )=0$) and neglect the
repulsion effects ($r<<a)$. Since the turning rate $\mathbb{T}_{\pm }$, Eq.\
(\ref{turn}), depends on both residence time $\tau $ and time $t$
(indirectly through $\rho _{+}$ and $\rho _{-}$), we can not define the
running time PDF. It can only be done for the linear case when $f=1$. For
this classical L\'{e}vy case the turning rate reads $\mathbb{T}(\tau )=\mu
/(\tau _{0}+\tau )$ and the running time PDF $\psi \left( \tau \right) $
defined in the standard way $\psi \left( \tau \right) =\mathbb{T}(\tau )\exp %
\left[ -\int_{0}^{\tau }\mathbb{T}(\tau )d\tau \right] $ \cite{Cox}, becomes
the power law density:
\begin{equation}
\psi (\tau )=\frac{\mu \tau _{0}^{\mu }}{(\tau _{0}+\tau )^{1+\mu }}.
\label{wtden}
\end{equation}%
For $1<\mu <2,$ this PDF has a finite first moment and infinite second
moment. This case corresponds to anomalous subballistic superdiffusion for
which the mean squared displacement is $\left\langle x^{2}\right\rangle \sim
t^{3-\mu }$ \cite{SWK,KBS,MK}.

Importantly, for individuals interacting via alignment we consider $f\neq
const.$ and $\mu >2$ for which the system without interactions has standard
long-time diffusive behaviour: $\left\langle x^{2}\right\rangle \sim t$ as $%
t\rightarrow \infty $ (curve $(1)$ in Fig.\ \ref{FIG2}). We do not assume
the anomalous running time PDF from the inception as its is done for a
classical theory of superdiffusive transport \cite{SWK,KBS,MK}. In our
simulations we chose $l_{a}=1$, $\alpha =1$ and $\beta =0$ in Eq.\ (\ref{A}%
), and an exponential interaction function $f(A_{\pm })=\exp (-A_{\pm })$.
For $\alpha =\beta =1$ we obtain similar results. At $t=0$ we consider a
uniform distribution of individuals in the interval $(-1,1)$.

\begin{figure}[t]
\vspace{-20pt} \centerline{ \psfig{figure=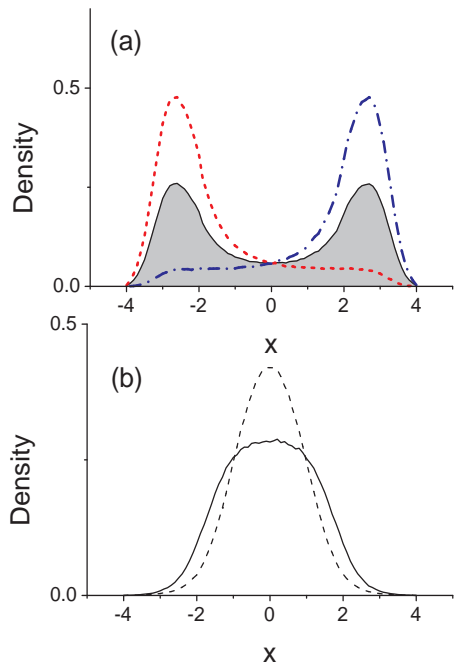,width=70mm,height=100mm} }
\caption{(a) Alignment leads to the formation of two moving aggregates known
as clumps. Density of the walkers $\protect\rho$ (solid curve) develops two
bumps corresponding to groups of walkers moving to the left $\protect\rho%
_{-} $ (dashed curve) and to the right $\protect\rho_{+}$ (dashed-dotted
curve). Here $a=2.5$ and other parameters are the same as in Fig.\ \protect
\ref{FIG2}. (b) For weak interactions, $a\rightarrow0$, we find no L\'{e}vy
walks and clumping behaviour. The solid curve corresponds to $a=0.1$.
Without interactions the density of walkers (dashed curve) is Gaussian apart
from tails. All densities were calculated at $t=3$.}
\label{FIG3}
\end{figure}

Figure \ref{FIG1} illustrates the emergence of the L\'{e}vy walk as the
ensemble averaged mean squared displacement (MSD) \cite{MSD} displays
superdiffusive behavior (curve (2)). For $\mu =3$ and $a=1.5$, we find $%
\left\langle \left( x-x_{0}\right) ^{2}\right\rangle \simeq t^{1.7}$ (other
parameters are listed in the figure caption). Since the individuals disperse
in space, their density and therefore the strength of interactions decrease
with time. That is, $f\rightarrow 1$ since $A_{\pm }\rightarrow 0$. As the
result, the walkers perform a normal diffusion at longer times (curve (2)).
Figure\ \ref{FIG2} confirms the emergence of the L\'{e}vy walk. It shows the
power law behaviour of the run length PDF with exponent $-2.7$ for the same
parameters used in Fig.\ \ref{FIG1}. A typical trajectory of an individual
involves long runs displaying anomalous persistence (walkers collectively
move in one direction) (Fig.\ \ref{FIG3} (b)). The results of our model
(Figs.\ \ref{FIG1}, \ref{FIG2}) qualitatively explain the emergence of the L%
\'{e}vy walk observed experimentally for swarming bacteria \cite{Main}. We
show that\ the standard switching (run-tumble) behaviour of individual is
drastically changed due to a collective motion that facilitates the L\'{e}vy
walk \cite{Main}. Interestingly, the alignment interactions lead\ to the
creation of two groups of individuals called clumps that move to the left
and to the right (Fig.\ \ref{FIG3} (a)). This is in agreement with the
clumping behaviour observed in non local hyperbolic models for
self-organized biological groups \cite{FE}. For small interaction strength $%
a\rightarrow 0$, there is no clumping phenomenon (Fig.\ \ref{FIG3} (b)).

\begin{figure}[t]
\vspace{-162pt}
\centerline{ \psfig{figure=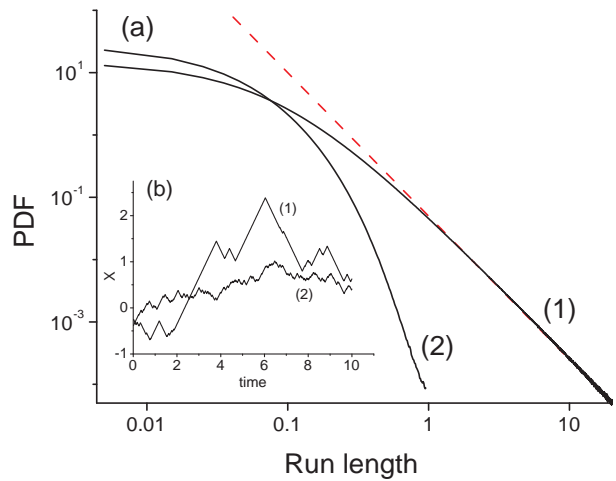,width=89mm,height=125mm}
}
\caption{(a) The run length PDF of individuals interacting repulsively.
Interactions lead to the transition to Brownian diffusion. Here we use $%
r=10^3$ and $\protect\mu=1.3$. Other parameters and simulation procedure are
the same. The run length PDF transitions from a power law with exponent $-%
\protect\mu-1 $ for the L\'evy walk without interactions (curve (1)) to
exponential distribution for the standard diffusion (curve (2)). (b) Typical
trajectories of walkers with interactions (curve (2)) and without
interactions (curve (1)). Repulsion interactions truncate the long runs of
the L\'{e}vy walks.}
\label{FIG4}
\end{figure}

\textit{Nonlinear transition from superdiffusion to diffusion.} We now
ignore the alignment effects ($f\left( A_{\pm }\right) =1$) and focus on the
repulsion/collision interactions. It follows from Eq.\ (\ref{turn}), the
switching rate $\mathbb{T}_{\pm }=\frac{\mu }{\tau _{0}+\tau }+\gamma _{\pm
}\left( \rho _{\mp }\right) $, where the interaction term $\gamma _{\pm }$
is defined in (\ref{col}). Note that now we consider the case $1<\mu <2$.
That is, without interactions we have a subballistic superdiffusive L\'{e}vy
walk with power law running time density Eq.\ (\ref{wtden}) and the MSD
growing as $\left\langle x^{2}\right\rangle \sim t^{3-\mu }$. We obtain
explicit expressions for the total turning rates $i_{\pm }$ in terms of the
density of walkers (for the derivation see the Supplementary Materials)
\begin{eqnarray}
i_{\pm }(x,t) &=&\int_{0}^{t}K(t-\tau )\rho _{\pm }(x\mp v(t-\tau ),\tau
)\times  \label{I+} \\
&&e^{-\int_{\tau }^{t}\gamma _{\pm }(\rho _{\mp }(x\mp v(t-u),u))du}d\tau .
\notag
\end{eqnarray}%
It is clear from (\ref{I+}) that the rate $\gamma _{\pm }$ plays the role of
a tempering parameter. This term is responsible for the shift of the
superdiffusive L\'{e}vy walk towards standard diffusion as the density $\rho
_{\pm }$ increases. The tempering effect of the repulsion/collision
interactions is similar to the tempering due to the random death of walkers
\cite{Fed4}. Figure \ref{FIG4} shows the results of numerical simulations
corresponding to the rate (\ref{turn}) with $f\left( A_{\pm }\right) =1$. In
the absence of repulsion we consider a superdiffusive L\'{e}vy walk with $%
\mu =1.3$. A typical trajectory (Fig.\ \ref{FIG4} (b)) has many long runs
and the distribution of the run length is a power law with exponent $-\mu -1$
(curve (1) in (a)). Repulsion/collision interactions drastically change the
stochastic dynamics of individuals. The long runs are truncated and the
trajectory appears Brownian (curve (2) in (b)). The run length PDF becomes
exponential (curve (2) in (a)), confirming the transition from a L\'{e}vy
walk to Brownian diffusion. Such a transition was observed experimentally in
the movement of mussels as their density increases \cite{MMM}.

\textit{Summary. }We have proposed a nonlinear persistent random walk model
of collectively moving individuals that interact via alignment and
repulsion. The walkers' interactions have been taken into account on the
mesoscopic level, at which the individuals' turning rate depends on the mean
field population density (nonlinear effect) and running time since the last
velocity switching (non-Markovian effect). The main result of this paper is
that the non-local alignment leads to the anomalous nonlinear persistence of
the random walkers and the emergence of the L\'{e}vy walk as a collective
phenomenon. Importantly this emergent superdiffusive movement of individuals
is a nonlinear non-Markovian effect, and is not based on the standard
assumption of a power-law running time distribution from the inception. We
should note that non-Markovian effects are crucial, since the numerical
simulations of the nonlinear model without the running time dependence in (%
\ref{turn}) show no L\'evy walks. We have qualitatively explained (1) the
experimentally observed emergence of superdiffusive L\'{e}vy walks of
swarming bacteria due to their collective dynamics and (2) the transition
from subballistic superdiffusion to the Brownian motion of individuals
interacting via repulsion/collision, which was observed in the movement of
mussels as their density increases. Our results are relevant to
experimentally observed superdiffusion of micron-scale beads in bacterial
bath \cite{Wu}, where it has been found that the superdiffusion occurs as a
result of the collective dynamics due to formations of coherent structures
like jets. Potentially, our model could be useful for studying collective
behavior of interacting individuals such as bacteria which use collective
movement for better protection against multiple antibiotics \cite{BWH}.

\textit{Acknowledgement}. This work was supported by EPSRC Grants No.
EP/J019526/1 and EP/N018060/1.

\newpage

\section{ Supplementary Materials}

\subsection{2-D generalization of nonlinear L\'{e}vy walk model}

In 2-D we consider the random motion of an individual that runs in the
direction $\mathbf{\theta }=(\cos \varphi ,\sin \varphi )$ with the constant
velocity $v$ \ during the running time $\tau $, and changes the direction at
$(\mathbf{x},t)$ to $\mathbf{\theta }^{\prime }=(\cos \varphi ^{\prime
},\sin \varphi ^{\prime })$. The turning rate $\mathbb{T}_{\rho }\left(
\mathbf{x},t,\tau ,\varphi ,\varphi ^{\prime }\right) $ from $\varphi $ to $%
\varphi ^{\prime }$ at $(\mathbf{x},t)$ depends on the running time $\tau $
and the non-local interactions with neighboring conspecifics. We define the
mean structural density of individuals, $n(\mathbf{x},t,\tau ,\varphi ),$ at
point $\mathbf{x}$ and time $t$ moving in the direction $\mathbf{\theta }$%
\textbf{\ }and having started the move a time $\tau $ ago\textbf{.} The
governing equation for $n(\mathbf{x},t,\tau ,\varphi )$ takes the form \cite%
{Alt}
\begin{equation}
\frac{\partial n}{\partial t}+v\mathbf{\theta \cdot \nabla }n+\frac{\partial
n}{\partial \tau }=-\gamma _{\rho }\left( \mathbf{x},t,\tau ,\varphi \right)
n,  \label{mainbal}
\end{equation}%
where $\gamma _{\rho }$ can be defined in terms of the turning rate $\mathbb{%
T}_{\rho }$ as follows
\begin{equation}
\gamma _{\rho }\left( \mathbf{x},t,\tau ,\varphi \right) =\int_{-\pi }^{\pi }%
\mathbb{T}_{\rho }\left( \mathbf{x},t,\tau ,\varphi ,\varphi ^{\prime
}\right) d\varphi ^{\prime }.  \label{gamma}
\end{equation}%
The function $\gamma _{\rho }\left( \mathbf{x},t,\tau ,\varphi \right) $
describes the rate at which the individual changes the direction at\ $(%
\mathbf{x},t)$ from $\varphi $ to other directions due to interactions with
neighboring conspecifics. We assume that at the initial time $t=0$ all
individuals have zero running time
\begin{equation}
n(\mathbf{x},0,\tau ,\varphi )=\rho (\mathbf{x},0,\varphi )\delta (\tau ).
\label{initial}
\end{equation}%
The total population density density is
\begin{equation}
\rho (\mathbf{x},t,\varphi )=\int_{0}^{t}n(\mathbf{x},t,\tau ,\varphi )d\tau
.  \label{den}
\end{equation}%
We set up the boundary condition at zero running time $\tau =0$:
\begin{equation}
n(\mathbf{x},t,0,\varphi )=\int_{0}^{t}\int_{-\pi }^{\pi }\mathbb{T}_{\rho
}\left( \mathbf{x},t,\tau ,\varphi ^{\prime },\varphi \right) n(\mathbf{x}%
,t,\tau ,\varphi ^{\prime })d\varphi ^{\prime }d\tau .  \label{in0}
\end{equation}%
From the Markovian equation (\ref{mainbal}) together with (\ref{initial})
and (\ref{in0}) one can\textbf{\ }obtain the equation for $\rho (\mathbf{x}%
,t,\varphi )$ \cite{Jake}
\begin{equation}
\frac{\partial \rho }{\partial t}+v\mathbf{\theta \cdot \nabla }\rho =-i(%
\mathbf{x},t,\varphi )+n(\mathbf{x},t,0,\varphi ),  \label{basic}
\end{equation}%
where%
\begin{equation}
i(\mathbf{x},t,\varphi )=\int_{0}^{t}\gamma _{\rho }\left( \mathbf{x},t,\tau
,\varphi \right) n(\mathbf{x},t,\tau ,\varphi )d\tau .
\end{equation}%
In the linear case without interactions, one can find $i(\mathbf{x}%
,t,\varphi )$ in terms of the total density $\rho (\mathbf{x},t,\varphi )$%
\cite{Jake}
\begin{equation}
i(\mathbf{x},t,\varphi )=\int_{0}^{t}K(t-\tau )\rho (\mathbf{x}-v\mathbf{%
\theta }(t-\tau ),\tau ,\varphi )d\tau ,  \label{i}
\end{equation}%
where $K(t)$ is the standard memory kernal \cite{Hel}.

Non-local interactions involving alignment rate $\mathbb{T}_{al}$ and
repulsion/collision rate $\mathbb{T}_{r}$ can be modelled as follows:%
\begin{equation}
\mathbb{T}_{\rho }=\mathbb{T}_{al}+\mathbb{T}_{r},
\end{equation}%
where
\begin{eqnarray}
\mathbb{T}_{al} &=&\frac{\mu }{\tau _{0}+\tau }\int_{-\pi }^{\pi }\int_{%
\mathbb{R}^{2}}K_{al}^{d}(\mathbf{x-y})K_{al}^{o}\left( \chi ,\varphi
^{\prime }\right) \times  \notag \\
&&\omega _{al}\left( \varphi ^{\prime }-\varphi ,\varphi ^{\prime }-\chi
\right) \rho (\mathbf{y},t,\chi )d\mathbf{y}d\chi ,
\end{eqnarray}%
\begin{eqnarray}
\mathbb{T}_{r} &=&r\left( \tau \right) \int_{-\pi }^{\pi }\int_{\mathbb{R}%
^{2}}K_{r}^{d}(\mathbf{x-y})K_{r}^{o}\left( \mathbf{x,y},\varphi ^{\prime
}\right) \times  \notag \\
&&\omega _{r}\left( \varphi ^{\prime }-\varphi ,\varphi ^{\prime }-\psi
\right) \rho (\mathbf{y},t,\chi )d\mathbf{y}d\chi .
\end{eqnarray}%
The explicit expressions for the functions $K_{al,r}^{d}$, $K_{al,r}^{o}$
and $\omega _{al,r}$ can be found in \cite{Fete,Car}. Detailed study of 2-D
model will follow.

\subsection{Equations for the unstructured densities $\protect\rho _{\pm
}(x,t)$}

Balance equations for the unstructured densities can be found by
differentiating
\begin{equation}
\rho _{\pm }(x,t)=\int_{0}^{t}n_{\pm }(x,t,\tau )d\tau  \label{ii3}
\end{equation}%
with respect to time $t.$ Because of the initial condition
\begin{equation}
n_{\pm }(x,0,\tau )=\frac{\rho (x,0)}{2}\delta (\tau ),  \label{iniini}
\end{equation}%
\ the running time $\tau $ varies from $0$ to $t$. We obtain for $\rho _{\pm
}$ the following equation
\begin{eqnarray*}
\frac{\partial \rho _{_{\pm }}}{\partial t} &=&n_{_{\pm }}(x,t,t)\mp
v\int_{0}^{t}\frac{\partial n_{_{\pm }}}{\partial x}d\tau -\int_{0}^{t}\frac{%
\partial n_{_{\pm }}}{\partial \tau }d\tau \\
&&-\int_{0}^{t}\mathbb{T}_{\pm }\left( \tau ,\rho _{+},\rho _{-}\right)
n_{\pm }d\tau .
\end{eqnarray*}%
Since a zero running time condition ($\tau =0)$ involves the proliferation
of the individuals with the proliferation rate $k\left( \rho \right) :$
\begin{equation}
n_{\pm }(x,t,0)=\int_{0}^{t}\left[ \mathbb{T}_{\mp }\left( \tau ,\rho
_{+},\rho _{-}\right) n_{\mp }+k\left( \rho \right) n_{\pm }\right] d\tau ,
\end{equation}%
we rewrite the equation for $\frac{\partial \rho _{_{\pm }}}{\partial t}$ as
follows
\begin{equation}
\frac{\partial \rho _{_{\pm }}}{\partial t}\pm v\frac{\partial \rho _{\pm }}{%
\partial x}=i_{_{\mp }}(x,t)-i_{_{\pm }}(x,t)+k\left( \rho \right) \rho
_{\pm },
\end{equation}%
where
\begin{equation}
i_{\pm }(x,t)=\int_{0}^{t}\mathbb{T}_{\pm }\left( \tau ,\rho _{+},\rho
_{-}\right) n_{\pm }(x,t,\tau )d\tau .  \label{ii1}
\end{equation}

\subsection{Nonlinear transition from superdiffusion to diffusion.}

Now let us find the switching rate $i_{\pm }(x,t)$ in terms of the density $%
\rho _{\pm }(x,t)$ for the rate%
\begin{equation}
\mathbb{T}_{\pm }\left( \tau ,\rho _{+},\rho _{-}\right) =\frac{\mu }{\tau
_{0}+\tau }+\gamma _{\pm }\left( \rho _{\mp }\right) .  \label{rrrr}
\end{equation}%
The purpose is show that $\gamma _{\pm }$ plays the role of nonlinear
tempering. By using the method of characteristics we solve the equation
\begin{equation}
\frac{\partial n_{\pm }}{\partial t}\pm v\frac{\partial n_{\pm }}{\partial x}%
+\frac{\partial n_{\pm }}{\partial \tau }=-\mathbb{T}_{\pm }\left( \tau
,\rho _{+},\rho _{-}\right) n_{\pm }.
\end{equation}%
We find for $\tau <t$
\begin{eqnarray}
n_{\pm }(x,t,\tau ) &=&n_{\pm }(x\mp v\tau ,t-\tau ,0)\times
\label{solution} \\
&&\Psi (\tau )e^{-\int_{t-\tau }^{t}\gamma _{\pm }(\rho _{\mp }(x\mp
v(t-u),u))du}.  \notag
\end{eqnarray}%
where the survival function $\Psi (\tau )$ is
\begin{equation}
\Psi (\tau )=\left( \frac{\tau _{0}}{\tau _{0}+\tau }\right) ^{\mu }.
\label{ss}
\end{equation}%
\ The formula (\ref{solution}) can be rewritten as
\begin{equation}
n_{_{\pm }}(x,t,\tau )=n_{_{\pm }}(x\mp v\tau ,t-\tau ,0)\Psi (\tau )\frac{%
F_{\rho }(x,t)}{F_{\rho }(x,\tau )}  \label{new5}
\end{equation}%
where
\begin{equation}
F_{\rho }(x,t)=e^{-\int_{0}^{t}\gamma _{\pm }(\rho _{\mp }(x\mp
v(t-u),u))du}.
\end{equation}%
Taking into account the initial condition (\ref{iniini}) and substituting (%
\ref{new5}) into we obtain
\begin{eqnarray}
i_{\pm }(x,t) &=&\int_{0}^{t}n_{_{\pm }}(x\mp v\tau ,t-\tau ,0)\psi (\tau )%
\frac{F_{\rho }(x,t)}{F_{\rho }(x,\tau )}d\tau  \notag \\
&&+\frac{1}{2}\rho (x\mp vt,0)\psi (t)F_{\rho }(x,t)  \label{jjj}
\end{eqnarray}%
and%
\begin{eqnarray}
\rho _{\pm }(x,t) &=&\int_{0}^{t}n_{_{\pm }}(x\mp v\tau ,t-\tau ,0)\Psi
(\tau )\frac{F_{\rho }(x,t)}{F_{\rho }(x,\tau )}d\tau  \notag \\
&&+\frac{1}{2}\Psi (t)\rho (x\mp vt,0)F_{\rho }(x,t).  \label{rr}
\end{eqnarray}%
where $\psi (\tau )=-d\Psi (\tau )/d\tau .$ By using the Laplace transforms
one can eliminate $n_{\pm }(x,t,0)$ from the above equations and obtain
explicit expressions for the total turning rates $i_{\pm }$ in terms of the
density of walkers $\rho _{\pm }:$
\begin{eqnarray}
i_{\pm }(x,t) &=&\int_{0}^{t}K(t-\tau )\rho _{\pm }(x\mp v(t-\tau ),\tau
)\times  \notag \\
&&e^{-\int_{\tau }^{t}\gamma _{\pm }(\rho _{\mp }(x\mp v(t-u),u))du}d\tau .
\label{rrr}
\end{eqnarray}%
This term involves the exponential factor in which the rate $\gamma _{\pm }$
plays the role of a tempering parameter. This rate leads to the shift of the
superdiffusive L\'{e}vy walk towards Brownian motion as the density $\rho
_{\pm }$ increases. The main feature of the rate $i_{\pm }(x,t)$ is that
although the rates $\frac{\mu }{\tau _{0}+\tau }$ and $\gamma _{\pm }\left(
\rho _{\mp }\right) $\ are additive (see (\ref{rrrr})), the corresponding
terms in the rate (\ref{rrr}) are not additive. This is clearly a
non-Markovian tempering effect.


\begin{thebibliography}{99}
\bibitem{MSD} The MSD is calculated in a standard way: $\left\langle \left(
x-x_{0}\right) ^{2}\right\rangle
=\sum\limits_{i=1}^{N}(x_{i}(t)-x_{i}(0))^{2}/N$, $x_{i}(t)$ is the position
of walker at time $t$ and $N=10^{6}$ is the number of walkers.
\end{thebibliography}
\end{document}